\documentclass[conference]{IEEEtran}
\IEEEoverridecommandlockouts
\usepackage{acronym}
\usepackage{cite}
\usepackage{amsmath,amssymb,amsfonts}
\usepackage[ruled]{algorithm2e}
\usepackage{textcomp}
\usepackage{xcolor}
\usepackage{lipsum}
\usepackage{cleveref}
\usepackage[load-configurations=binary, binary-units=true]{siunitx}
\usepackage[a4paper, top=19mm, bottom=43mm, left=13mm, right=13mm]{geometry}

\DeclareSIUnit{\belmilliwatt}{Bm}
\DeclareSIUnit{\dBm}{\deci\belmilliwatt}
\DeclareSIUnit{\nothing}{\relax}

\usepackage{booktabs}
\ifCLASSOPTIONcompsoc
\usepackage[caption=false,font=normalsize,labelfont=sf,textfont=sf]{subfig}
\else
\usepackage[caption=false,font=footnotesize]{subfig}
\fi
\usepackage{graphicx}
\usepackage{pgf,tikz}
\usepackage{pgfplotstable}
\usetikzlibrary{pgfplots.statistics, pgfplots.colorbrewer} 
\usetikzlibrary{positioning,arrows.meta,patterns,backgrounds,fit}
\definecolor{chart1}{HTML}{CA472F} % red        CA472F - [0.79216 0.27843 0.18431]
\definecolor{chart2}{HTML}{9DD866} % green      9DD866 - [0.61569 0.84706 0.40000]
\definecolor{chart3}{HTML}{6F4E7C} % purple     6F4E7C - [0.43529 0.30588 0.48627]
\definecolor{chart4}{HTML}{F6C85F} % yellow     F6C85F - [0.96471 0.78431 0.37255]

\acrodef{BLE}{Bluetooth Low Energy}
\acrodef{AoA}{Angle-of-Arrival}
\acrodef{RSSI}{Received Signal Strength Indication}
\acrodef{RMSE}{Root mean square error}
\acrodef{CTE}{Constant Tone Extension}
\acrodef{IoT}{Internet-of-Things}
\acrodef{TDoA}{Time-Difference-of-Arrival}
\acrodef{MUSIC}{Multiple Signal Classification}
\acrodef{UWB}{Ultra-Wide-Band}
\acrodef{RF}{Radio Frequency}
\acrodef{PCB}{Printed Circuit Board}
\acrodef{I/Q}{In-Phase/Quadrature}
\acrodef{SDR}{Software Defined Radio}
\acrodef{SoC}{System-on-Chip}

\newcommand{\nota}[1]{#1}

\let\olip\lipsum
\renewcommand{\lipsum}[1][]{\nota{\olip[#1]}}

\begin{document}

\title{Design and Experimental Evaluation of a Bluetooth 5.1 Antenna Array for Angle-of-Arrival Estimation\\
%\thanks{This work is financed by the ERDF – European Regional Development Fund through the Operational Programme for Competitiveness and Internationalisation - COMPETE 2020 Programme within project SLID: Stock Live IDentification (POCI-01-0247-FEDER-045388).}

% decarbonize acks
\thanks{This work is financed by the ERDF - European Regional Development Fund through the Norte Portugal Regional Operational Programme - NORTE 2020 under the Portugal 2020 Partnership Agreement within project DECARBONIZE, with reference NORTE-01-0145-FEDER-000065.}
}

% strx Acks
%\thanks{\nota{This work is supported by the European Regional Development Fund (FEDER), through the Competitiveness and Internationalization Operational Programme (COMPETE 2020) of the Portugal 2020 framework  within Project STRx with Nr. 033623 (POCI-01-0247-FEDER-033623).}}

\author{
\IEEEauthorblockN{Nuno Paulino and Luís M. Pessoa}
\IEEEauthorblockA{INESC TEC and Faculty of Engineering,\\University of Porto, Porto, PORTUGAL\\ \{nuno.m.paulino, luis.m.pessoa\}@inesctec.pt}
\and
\IEEEauthorblockN{André Branquinho and Edgar Gonçalves}
\IEEEauthorblockA{\textit{Wavecom -- Soluções Rádio SA},\\ Aveiro, PORTUGAL \\
\{abranquinho, egoncalves\}@wavecom.pt}
}
% make the title area
\maketitle

\begin{abstract}
One the of the applications in the realm of the \ac{IoT} is real-time localization of assets in specific application environments where satellite based global positioning is unviable. Numerous approaches for localization relying on wireless sensor mesh systems have been evaluated, but the recent \ac{BLE} 5.1 direction finding features based on \ac{AoA} promise a low-cost solution for this application. In this paper, we present an implementation of a \ac{BLE} 5.1 based circular antenna array, and perform two experimental evaluations over the quality of the retrieved data sampled from the array.
\nota{Specifically, we retrieve samples of the phase value of the Constant Tone Extension which enables the direction finding functionalities through calculation of phase differences between antenna pairs.} 
\nota{We evaluate the quality of the sampled phase data} in an anechoic chamber, and in a real-world environment using a setup composed of four \ac{BLE} beacons. 
\end{abstract}


%%%%%%%%%%%%%%%%%%%%%%%%%%%%%%%%%%%%%%%%%%%%%%%%%%%%%%%%%%%%%%%%%%%%%%%%%%%%%%
\section{Introduction}

The push towards the long promise of the \ac{IoT} is gaining momentum as more efficient telecommunications are implemented, concurrently with increasingly more feature-full \ac{SoC} solutions. Together, these advances are powering more ambitious edge computing applications. One such application is real-time localization of devices or assets in context-specific indoor spaces \cite{8692423}. 
%
% ... such applications impose data processing requirements on these devices, such that real-time operation of the desired features is possible. 

Towards this, direction finding capabilities have been recently added to the Bluetooth Specification, as of version 5.1 \cite{ble51}. Specifically, direction finding is performed by sampling a \ac{CTE} which is appended to the end of a Bluetooth packet. A \ac{RF} device with an array of antennas, can sample the phase of this tone at each of the antennas, and derive the \ac{AoA} of the signal via the phase differences between antennas. Specific details of the processing depend on the number of antennas and their physical arrangement. 

% TODO: mention BLE, but end the intro on a twist into what it implies regarding signal processing
% Thus process entails data sampling and processing, which, should ideally be fully 

In this paper, we consider an application context where mobile \ac{BLE} receivers are equipped with an antenna array, envisioning application scenarios for self-localization of vehicles in warehouses, shipping ports, or similar environments. In these scenarios, the number of receivers wishing to know their location is known, and potentially small. So this allows for the installation of fixed, cheap \ac{BLE} beacons in legacy locations which may not have the wall-power and ethernet infrastructure required by the opposite solution, i.e., when the arrays are fixed and numerous, and the number of assets to locate is unknown and have only one antenna (e.g., exhibition halls or shopping malls). 

In previous work, we performed a simulation based evaluation for positioning and tracking \cite{9482525}, based on this topology, as a function of the number of beacons, receiver movement, among others. As the receiver moves, fast computation is required, as well as a reduced number of packets per fixed transmitter, to compute a position. We demonstrated that \ac{AoA} data received by multiple transmitters can allow the receiver to compute its own position. However, computing the \ac{AoA} requires prior signal processing over the previously mentioned phase samples. % meaning that the \ac{AoA} is dependant on sample quality and 
Therefore, if possible, computational complexity should be reduced in order to achive this in real-time on such edge devices.
    % ( is too computationally complex? see one of the refs in the related work section about this)
    
In this paper, we focus on obtaining and processing the \ac{CTE} phase samples, and evaluating the quality of the obtained data in a real-world scenario, by designing and implementing a \ac{PCB} with a circular antenna array with 8 antennas. We present the theoretical expected behaviour using generated phase data, verify the correct operation of the design in an anechoic chamber, and finally evaluate the quality of the data attainable in an outdoor environment free of obstructions. 

% TODO: put this where? its important kinda...
% We compute the phase difference data for our physical antenna arrangement, which is required to subsequently extract an estimate of the \ac{AoA}, but this final step is out of the scope of this paper.

%%%%%%%%%%%%%%%%%%%%%%%%%%%%%%%%%%%%%%%%%%%%%%%%%%%%%%%%%%%%%%%%%%%%%%%%%%%%%%
\section{Related Work}
\label{sec:rwork}

% 2022
% https://ieeexplore.ieee.org/document/9674235
Two data processing steps are proposed in \cite{9674235}, in order to improve \ac{AoA} estimations from raw phase data, including a comparison with the state-of-the-art algorithm \ac{MUSIC} \cite{1143830}. Using one commercial transmitter and one commercial linear array, the authors retrieve 200 packets per orientation between the two devices in steps of \SI{10}{\degree} in the range of \SI{-90}{\degree} to \SI{90}{\degree} (the range the linear receiver is capable of disambiguating). The devices are \SI{1}{\meter} off the ground and \SI{2}{\meter} apart in an indoor location. The first data processing step, a non-linear recursive least square method, intends to mitigate the effect of multi-path and noise. A second step employs an unscented Kalman filter to reduce the effect of different oscillator frequencies between transmitter and receiver (introduced to account for manufacture variation). For the described setup, the pre-processing leads to a decrease in the computed \ac{AoA} of \SI{3.8}{\degree} on average, although at the cost of increased computational load. 
% TODO: flaws and differences to us

% 2020
% https://ieeexplore.ieee.org/document/9190573
% NOTE: this paper uses exactly the same pre-processing steps as the previous one, apparently
Similarly, comparable pre-processing steps are applied to the raw \ac{I/Q} samples in \cite{9190573}. Namely, a non-linear least square curve fitting method to reduce the effect of noise on the \ac{I/Q} samples, a Kalman filter to address frequency and phase offsets inherent to the different antennas in the array and the switching process. A third step is added relative to \cite{9674235}, where a Gaussian filter compensates for estimation errors which are introduced depending on which \ac{BLE} data channel is being used, as the approach assumes a connected mode. For one commercial transmitter and receiver pair, with a linear array with 3 elements, placed \SI{1}{\meter} apart in an unspecified test environment, the approach reduces the estimation error significantly in the range of \SI{-60}{\degree} to \SI{60}{\degree}. As with \cite{9674235}, \ac{AoA} estimation for linear arrays significantly decays when the incident angle is parallel to the array. 
% TODO: flaws and differences to us
    % TODO: the authors demonstrate this for one received packet, but do not specify how many packets were processed via the approach, 

% 2021
% https://ieeexplore.ieee.org/abstract/document/9369638
In \cite{9369638} the \ac{BLE} 5.1 direction-finding is implemented via \ac{SDR} on a Xilinx Zynq-7000 System-on-Chip. The authors design their own linear array with 4 rectangular patch microstrip antennas, and employ a commercial radio module and \ac{RF} switch. A single \ac{BLE} beacon is placed \SI{3}{\meter} away from the receiver setup, and the beacon is moved along the azimuth plane (parallel to the forward face of the antenna array), for a range of \SI{-90}{\degree} to \SI{90}{\degree}. Resorting to the \ac{MUSIC} algorithm, the design achieves a root mean square error up to \SI{5}{\degree}. 

% TODO: good 2021 ref on how circular anteanna arrays are still being studied
% https://link.springer.com/article/10.1007/s00500-021-05778-2 ???

%The required \ac{BLE} packet characteristics for compliance with the 5.1 specification for direction-finding are implemented 

% 2021
% https://www.mdpi.com/2072-4292/13/21/4301/htm
%In \cite{rs13214301}, 

% https://thescipub.com/pdf/ajassp.2012.1979.1984.pdf

% spherical array, 2022
% https://www.mdpi.com/2079-9292/11/2/208/pdf

% [15] M. Kulin, T. Kazaz, I. Moerman, and E. De Poorter, “End-to-End
 %Learning From Spectrum Data: A Deep Learning Approach for Wire less Signal Identifification in %Spectrum Monitoring Applications,” IEEE
 %Access, vol. 6, pp. 18484-18501, March 2018
    % TODO: cite this for an pproach which uses AI to classify the data? must read

%%%%%%%%%%%%%%%%%%%%%%%%%%%%%%%%%%%%%%%%%%%%%%%%%%%%%%%%%%%%%%%%%%%%%%%%%%%%%%
\section{Proposed Approach}

\nota{Regarding the design of the antenna array, as \Cref{sec:rwork} mentions, arrays are typically either linear or circular. Linear arrays are easier to implement, but suffer from ambiguity regarding the true \ac{AoA}, while circular arrays offer data redundancy due to inherent symmetry \cite{USHAPADMINI1994243}. Therefore, although circular array designs are not novel, they are still actively used and studied for direction finding applications \cite{9166111,10.1007/s00500-021-05778-2}. Due to the recent direction finding features added to \ac{BLE}, they are now being explored for use in combination with this protocol.}

In this paper, we evaluate a \ac{BLE} based localization approach using our own design and software for a circular antenna array with 8 antennas. The following sections explain the physical design of the board, the signal processing involved, and the theoretical expected behaviour.

%%%%%%%%%%%%%%%%%%%%%%%%%%%%%%%%%%%%%%%%%%%%%%%%%%%%%%%%%%%%%%%%%%%%%%%%%%%%%%
\subsection{Design of Circular 8-Antenna Array}

We have designed and fabricated a circular printed circuit board equipped with a single Nordic Semiconductor nRF52811 micro-controller \cite{nordic1}. This micro-controller supports the direction finding features specified by \ac{BLE} 5.1 \cite{ble51}.

The board is \SI{13}{\centi\meter} in diameter, and its 8 antennas are equally spaced, i.e., placed at \SI{45}{\degree} steps relative to the center of the board. At the center, the board contains at SKY13418-485LF switch \cite{sky1}, which is a SP8T \ac{RF} switch rated up to \SI{6}{\giga\hertz}. The tracks from the switch to each SMA connector have equal lengths. 
Besides these components, the board contains a 10-pin header, used for programming the micro-controller via an external devkit \cite{nordic2}, via a J-Link interface, and also to serve as the serial interface for data transfer from the antenna array board to a workstation. A micro-USB connector is used only for power, with a \SI{5}{\volt} to \SI{2.5}{\volt} regulator \cite{tps}. This regulator supplies both the radio switch, and the micro-controller. The board was fabricated in an FR-4 substrate with a thickness of \SI{1}{\milli\meter}.

The micro-controller is programmed in \emph{C}, and we rely on Nordic's own software development kit to configure the radio parameters for both the receiver and transmitter software versions. On the receiver, control of the \ac{RF} switch is done via three control pins between the micro-controller and the switch. During reception of a packet, the micro-controller's follows a preset switching pattern (configured at boot), which repeats, to select one antenna per switching period. 
During sampling, a data array in memory is directly filled with all received samples by the nrRF52811's integrated radio. The radio is capable of providing the phase samples both in \ac{I/Q} format or magnitude-phase format. We employ the latter format, and process the phase values given directly by the radio. Each sample is represented in a $[-201, 201]$ fixed-point value range, stored in a \SI{16}{\bit} integer. We currently transmit all phase samples to a workstation and compute the phase differences, and do not currently use the magnitude value.

% diagram of board
\begin{figure}
\centering
\includegraphics[width=0.8\linewidth]{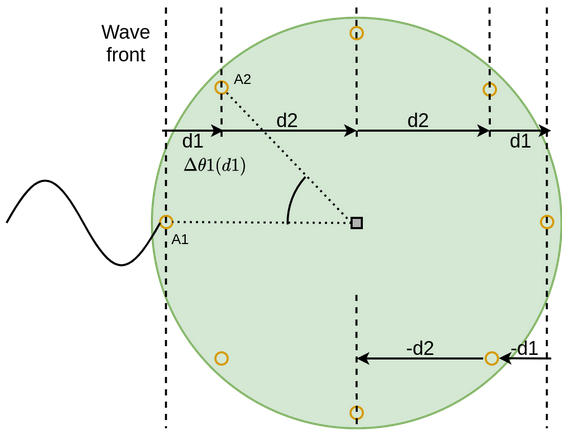}
\caption{Model of circular antenna array, and relationship between angle, and travelled distance between antenna pairs}
\label{fig:model}
\end{figure}

\Cref{fig:model} illustrates the simplified board design (tracks omitted), and how sampling multiple antennas in the array, in a known pattern, allows for determination of the incident \ac{AoA} by the receiver itself in a \SI{360}{\degree} range. Antennas are numbered sequentially clockwise. As with linear arrays, the distance travelled by the incident electromagnetic plane wave front between neighbouring antennas depends on the incident angle. 
For a circular array, this implies a symmetry in the phase differences between certain antenna pairs. For example, the same phase differences will be observed between antennas A2 and A3, and A3 and A4, as the same distance \emph{d2} is travelled by the wave front, as shown in the following section. 

%%%%%%%%%%%%%%%%%%%%%%%%%%%%%%%%%%%%%%%%%%%%%%%%%%%%%%%%%%%%%%%%%%%%%%%%%%%%%%
\subsection{Phase Difference Calculation}

The \ac{BLE} 5.1 specification for a direction-finding enabled packet states that the \ac{CTE} portion begins with a \SI{4}{\micro\second} guard period, followed by a \SI{8}{\micro\second} reference period, during which one antenna set as reference is sampled at a higher rate, before the switching pattern begins. The reference samples can be used to predict the phase values %of that antenna 
at subsequent timesteps, during which the following antennas are sampled \cite{cominelli2019dead}.

However, we do not utilize the reference samples, and directly compute the phase differences by using the samples retrieved in sequence, from two neighbour antennas. As our switching period is set to \SI{4}{\micro\second}, and since the \SI{250}{\kilo\hertz} frequency of the \ac{CTE} implies that the phase varies 90º per \SI{1}{\micro\second}, then each \emph{i-th} phase sample from an antenna will have an equivalent value after \SI{4}{\micro\second}. 

Therefore, we take all the samples from all antennas, and first apply an unwrapping step to the phase values, shown in \Cref{alg1}. This is required since, although the phase progresses linearly, its reported value is bound to the $[-201, 201]$ range given by the nRF52811's radio. We convert this fixed-point representation to floating-point within the more easily interpretable $[-180, 180]$ range.

% TODO: candidate for removal if lack of space

\begin{algorithm}
$nrSampleGroups\gets 32$\;
$nrSamples\gets 3$\;
\For{$i < nrSampleGroups$}{
    \For{$j < nrSamples$} {
        \If{$sample_{ij} < sample_{ij-1} - 180$}{
            $k\gets j$\;
            \For{$k < nrSamples$} {
                $sample_{ik} = sample_{ik} + 360;$
            }
        }
    }
}
\caption{Algorithm to unwrap the \emph{j} phase samples from each \emph{i-th} antenna from the $[-180, 180]$ bound}
\label{alg1}
\end{algorithm}
%
% for i < nrSampleGroups
%  for j < nrSamples
%   if(sample_ij < sample_ij-1 - 180)
%      for(k = j, j < 3)
%        sample_ik += 360;
%
% \caption{in our case, nrSampleGroups = 32, and nrSamples = 3, when using data sampled from the implemented design, but the same is applicable for generated data where nrSampleGroups and nrSamples take any values}

To compute each of the $n = 0,...,N$  phase differences, we take the $i = 0,...,S$ samples from each pair of neighbour antennas, subtract each 
%To compute each of the 32 phase differences, 
%We then subtract each 
\emph{i-th} sample pair, and take the mean of the differences as the phase difference for that antenna pair. Finally, we compute the modulus to bind the value to a \SI{-180}{\degree} to \SI{180}{\degree} range, as per \Cref{alg2}.
%
% explain that we are discarding 5 out of the 8 samples, since we previously determiend in anechoic chamber that those were bad
    % this reduces the uart tx time from reciver to laptop and speds up the experiments
    % TODO: how long did it take to take all packets at all locations? check logs of packets!
In this case $N=32$, and $S=2$. We utilized only 3 out of the 8 samples per antenna as a result of previous experiments in the anechoic chamber (omitted for brevity), where we observed that 5 out of the 8 samples  were too unreliable. We retain the second to fourth samples, as the remaining are affected by the switching behaviour of the \ac{RF} switch. 

\SetKwComment{Comment}{//}{}
\begin{algorithm}
%$i \gets 1$\;
$diffs[32] \gets 0;$ \Comment{32 phase differences}
$nrSampleGroups\gets 32$\;
$nrSamples\gets 3$\;
\For{$i < nrSampleGroups$}{
    %$j \gets 0$\;
    $d[nrSamples] \gets 0$\;
    \For{$j < nrSamples$}{
        $d[j] = sample_{i-1,j} - sample_{i,j}$\;
    }
    $diffs[i] = mod(mean(d) + 180, 360) - 180$\;
}
\caption{Calculation of phase difference between antenna pairs}
\label{alg2}
\end{algorithm}

% TODO: explain the signal processing portion here?
% TODO: how many subtractions, multiplications, divisions, etc, per packet??

% In terms of embedded computing, the approach requires that all the sampled phase data be processed efficiently...

% The implemented system requires significant signal processing in order to extract the final designed \ac{AoA} from the sampled phases data of the \ac{CTE}. Specifically, 

%%%%%%%%%%%%%%%%%%%%%%%%%%%%%%%%%%%%%%%%%%%%%%%%%%%%%%%%%%%%%%%%%%%%%%%%%%%%%%
\subsection{Expected Behaviour} % TODO: remove this subsection and merge with previous?

Knowing the geometry of the board, the physical placement of its antennas, the frequency of the \ac{CTE}, and the sampling parameters, we can generate simulated phase sample data. The purpose of this is to validate the algorithms for phase difference calculation in an ideal case, unaffected by potential design flaws of the board, or ambient conditions, which would compromise any further calculations. 
\noindent\begin{figure}%
    \centering%
        \begin{tikzpicture}%
        \begin{axis}[%
        xticklabels={
            $\Delta\theta 1$, 
            $\Delta\theta 2$, 
            $\Delta\theta 3$, 
            $\Delta\theta 4$, 
            $\Delta\theta 5$, 
            $\Delta\theta 6$, 
            $\Delta\theta 7$, 
            $\Delta\theta 8$},
        xtick=data,
        height=0.5\linewidth,
        width=1\linewidth,
        grid = both,
        major grid style = {lightgray},
        minor grid style = {lightgray!25},
        xlabel=Antenna Pair,
        ylabel=Phase Difference,
        legend columns = 2,
        legend style={
            fill=none, 
            draw=none, 
            at={(0.3,1.2)}, 
            anchor=north,
            /tikz/every even column/.append style={column sep=0.19cm}
        }]

        \addplot[chart1, thick, mark=x] 
        table [x=pair, y=A0, col sep=comma] {./data/theory1.csv};
        
        \addplot[chart2, thick, mark=x] 
        table [x=pair, y=A75, col sep=comma] {./data/theory1.csv};

        \addlegendentry{AoA = 0º}
        \addlegendentry{AoA = 75º}
        \end{axis}
        \end{tikzpicture}
\caption{Profile of calculated phase difference values for all adjacent antenna pairs, as a function of two different AoAs, using phase sample data generated according to the mode (\nota{e.g., pair 1 is the difference between the phase value between A1 and A2 ($\Delta\theta 1$), at the same instant in time)}}
\label{fig:theory1}
\end{figure}
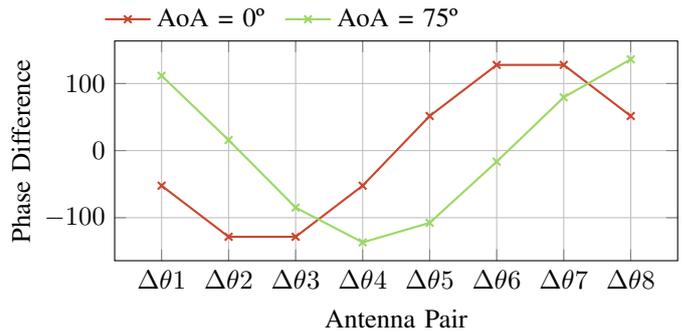

% TODO: candidate for removal?? also theory2??

\Cref{fig:theory1} illustrates the eight phase differences, for the eight antenna pairs, generated by processing generated phase data for two incident angles (as data is generated, the exact \ac{AoA} is illustrative and not related to the board design). Due to the board's geometry, the resulting profile of the phase differences is itself a sinusoidal wave, whose own phase shifts as a function of the \ac{AoA}. Therefore, determining this phase, e.g., via curve fitting, would be a subsequent step to arrive at an \ac{AoA}. 

% effects of noise
    % put plots affected by noise generated by matlab
\noindent\begin{figure}%
    \centering%
        \begin{tikzpicture}%
        \begin{axis}[%
        xticklabels={
            $\Delta\theta 1$, 
            $\Delta\theta 2$, 
            $\Delta\theta 3$, 
            $\Delta\theta 4$, 
            $\Delta\theta 5$, 
            $\Delta\theta 6$, 
            $\Delta\theta 7$, 
            $\Delta\theta 8$},
        xtick=data,
        height=0.5\linewidth,
        width=1\linewidth,
        grid = both,
        major grid style = {lightgray},
        minor grid style = {lightgray!25},
        xlabel=Antenna Pair,
        ylabel=Phase Difference,
        legend columns = 4,
        legend style={
            fill=none, 
            draw=none, 
            at={(0.45,1.2)}, 
            anchor=north,
            /tikz/every even column/.append style={column sep=0.19cm}
        }]

        \addplot[chart1, thick, mark=x] 
        table [x=d, y=e0, col sep=comma] {./data/effectOfNoise.csv};
        \addlegendentry{$\sigma$ = 0º}
        
        \addplot[chart2, mark=x] 
        table [x=d, y=e45, col sep=comma] {./data/effectOfNoise.csv};
        \addlegendentry{$\sigma$ = 45º}
        
        \addplot[chart3, mark=x] 
        table [x=d, y=e55, col sep=comma] {./data/effectOfNoise.csv};
        \addlegendentry{$\sigma$ = 55º}
        
        \addplot[chart4, mark=x] 
        table [x=d, y=e65, col sep=comma] {./data/effectOfNoise.csv};
        \addlegendentry{$\sigma$ = 65º}
        
        \end{axis}
        \end{tikzpicture}
\caption{Effect of guassian error with zero mean ($\mu = 0$) and for different standard deviations ($\sigma$)}
\label{fig:theory2}
\end{figure}
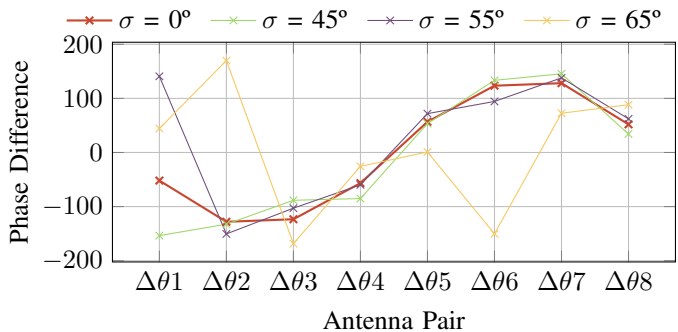
The resulting phase difference profile is in practice affected by noise, which originates from multi-path effects, differences in oscillators between transmiters and receivers, behaviour of the \ac{RF} switch, or even slight variations in impedances of the antenna tracks or physical separation between antenna pairs. To illustrate this, we introduce gaussian noise to the phase samples we generate via simulation, and \Cref{fig:theory2} demonstrates the degradation of the profile based on noise with zero mean, and the shown standard deviation ($\sigma$).

% (Therefore, multiple subsequent steps can be conceived to...)

%%%%%%%%%%%%%%%%%%%%%%%%%%%%%%%%%%%%%%%%%%%%%%%%%%%%%%%%%%%%%%%%%%%%%%%%%%%%%%
\subsection{Position Calculation from Multiple Angles-of-Arrival}

\nota{In this paper, we do not yet advance into location calculation, %both for reasons of brevity, and as the experimental evaluation in \Cref{sec:exp} will show, the limitations imposed by the quality of the attained data. 
and focus on the quality of the attained phase data in the presented experimental environment. However, we will shortly explain the method we previously validated in \cite{9482525}.}

\nota{We target application scenarios where a mobile receiver wishes to self locate. Therefore, by relying on knowledge of the map, and on the absolute position fixed position of the beacons, multiple \acp{AoA} would generate vectors that, in ideal conditions, would intersect in a single point. In such a case, even two \acp{AoA} would be sufficient for receiver localization. However, due to noise present in the \ac{CTE} phase samples, an intersection between two vectors alone could lead to a significant positioning error. Position calculation improves with more \acp{AoA} from different source beacons. Even so, we can only compute a candidate area, i.e., polygon, for the position of the receiver. We thus resort to state-of-the-art least-squares method which computes a point whose total distance to each side of the polygon is smallest \cite{traa2013least,9482525}.}

% TODO: some comment about future work etc
% TODO: could be useful as fast citation when mentioning future work using NN?
% A. Khan, S. Wang and Z. Zhu, “Angle-of-Arrival Estimation Using
 %an Adaptive Machine Learning Framework,” in IEEE Communications
 %Letters, vol. 23, no. 2, pp. 294-297, Feb. 2019.
 
%\nota{In other recent works, machine learning approaches are being applied to estimation of the \ac{AoA} \cite{8554304,9336252}. We are also }

% "The traditional spectral- or parametric-based AOA estimation methods are difficult to obtain real-time AOA information because of the relatively high computational complexity. " from one of these NN papers
 
%%%%%%%%%%%%%%%%%%%%%%%%%%%%%%%%%%%%%%%%%%%%%%%%%%%%%%%%%%%%%%%%%%%%%%%%%%%%%%
\section{Experimental Evaluation}
\label{sec:exp}

Using a total of five boards, we conducted two experimental evaluations. In both cases, one board was used as a receiver, while the remaining four were programmed as transmitters. The transmitter code, past the configuration portion, is a simple loop which broadcasts a direction-finding enabled packet where the payload is the transmitter's unique ID. The transmission power was set to \SI{4}{\dBm}. 

For all tests, we configured the \ac{CTE} length to \SI{160}{\micro\second}, the switching period to \SI{4}{\micro\second}, and the sample rate to \SI{500}{\nano\second}. This means that each \ac{BLE} packet produces 304 samples of it's constant tone's phase, 8 per sampled antenna. The sampling pattern is circular, meaning sampling starts from antenna 1, and follows the board perimeter. Since the switching pattern repeats, the full perimeter of the board is sampled multiple times during the \ac{CTE}. Specifically, four sampling rotations are performed, leading to 32 phase differences. 

Firstly, we validated the expected behaviour of the receiver in an anechoic chamber, and secondly, we evaluated the quality of the received data in a field test. 

%%%%%%%%%%%%%%%%%%%%%%%%%%%%%%%%%%%%%%%%%%%%%%%%%%%%%%%%%%%%%%%%%%%%%%%%%%%%%%
% anechoic chamber
\subsection{Phase Data Gathering in Anechoic Chamber}

\begin{figure}
\centering
\includegraphics[width=0.8\linewidth]{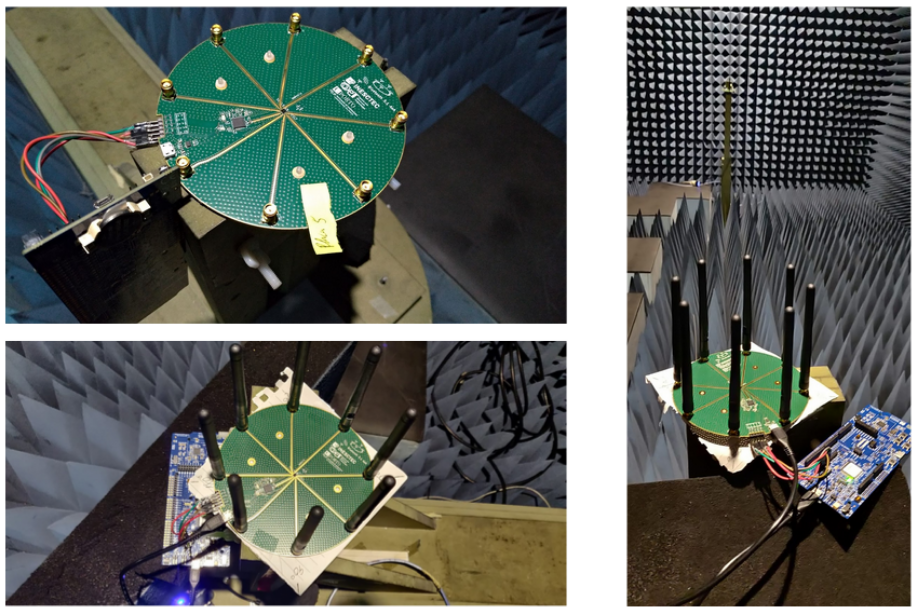}
\caption{Fabricated board design, and setup for retrieval of packets in anechoic chamber}
\label{fig:chamber}
\end{figure}

\Cref{fig:chamber} shows an experimental setup in an anechoic chamber. A single transmitter is placed in a static support, while the receiver is placed in a rotating mount. We rotate the receiver in steps of \SI{15}{\degree}, and gather one hundred packets per orientation. The distance between the transmitter and receiver is approximately \SI{5}{\meter}. 

For all retrieved packets per orientation, we computed the respective phase difference profiles, using the total of 104 samples retrieved per packet, arriving at the 32 differences between neighbouring antennas. \Cref{fig:chamberresults} illustrates this for 100 packets and 3 different orientations, demonstrating that the sampled phase data and phase differences follow the expected behaviour. 

\input{tikz/anechoicChamber}

%%%%%%%%%%%%%%%%%%%%%%%%%%%%%%%%%%%%%%%%%%%%%%%%%%%%%%%%%%%%%%%%%%%%%%%%%%%%%%
\subsection{Phase Data Gathering in Field Test}

A field test was conducted with four boards programmed as transmitters, with only one antenna, and one board programmed as receiver, making use of the entire 8-antenna array. 

\begin{figure*}
    \centering
    \subfloat[Aerial view of the field test location: an unobstructed space with a rugged concrete floor at the Faculty of Engineering of the University of Porto 
    %(N41º10'40, W8º35'40)
    \label{fig:location}]{
    \begin{tikzpicture}
        \node[anchor=south west,inner sep=0] at (0,0) {\includegraphics[width=0.35\linewidth]{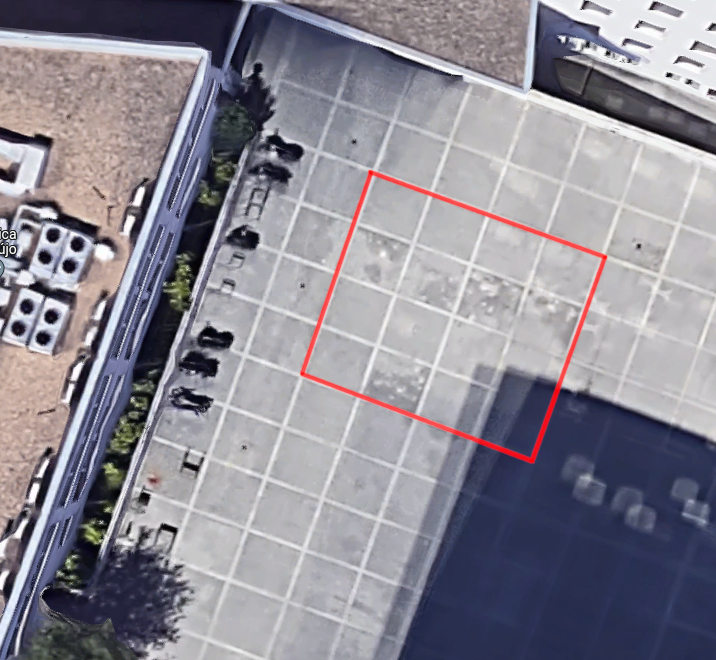}};
        %
        %\draw[red,ultra thick,rounded corners] (7.5,5.3) rectangle (9.4,6.2);
        \node[fill=white, align=center,text width=1cm,rounded corners] 
        at (3.8,5.0) {b2 at (0,0)};
       
        \node[fill=white, align=center,text width=1cm,rounded corners] 
        at (2.3,2.0) {b4 at (0,4)};
       
        \node[fill=white, align=center,text width=1cm,rounded corners] 
        at (5.5,1.3) {b1 at (4,4)};
     
        \node[fill=white, align=center,text width=1cm,rounded corners] 
        at (5.6,4.3) {b5 at (4,0)};
        
        \draw[red] (3.0,3.45) -- (5.05,2.7) node[midway, above] {12m};
    \end{tikzpicture}
    }
    \qquad
    \subfloat[Receiver placed along one of the edges of the outlined space, with laptop arrangement to store received \ac{BLE} packet phase data from all four transmitters.\label{fig:groundlocation}]{\includegraphics[width=0.35\linewidth]{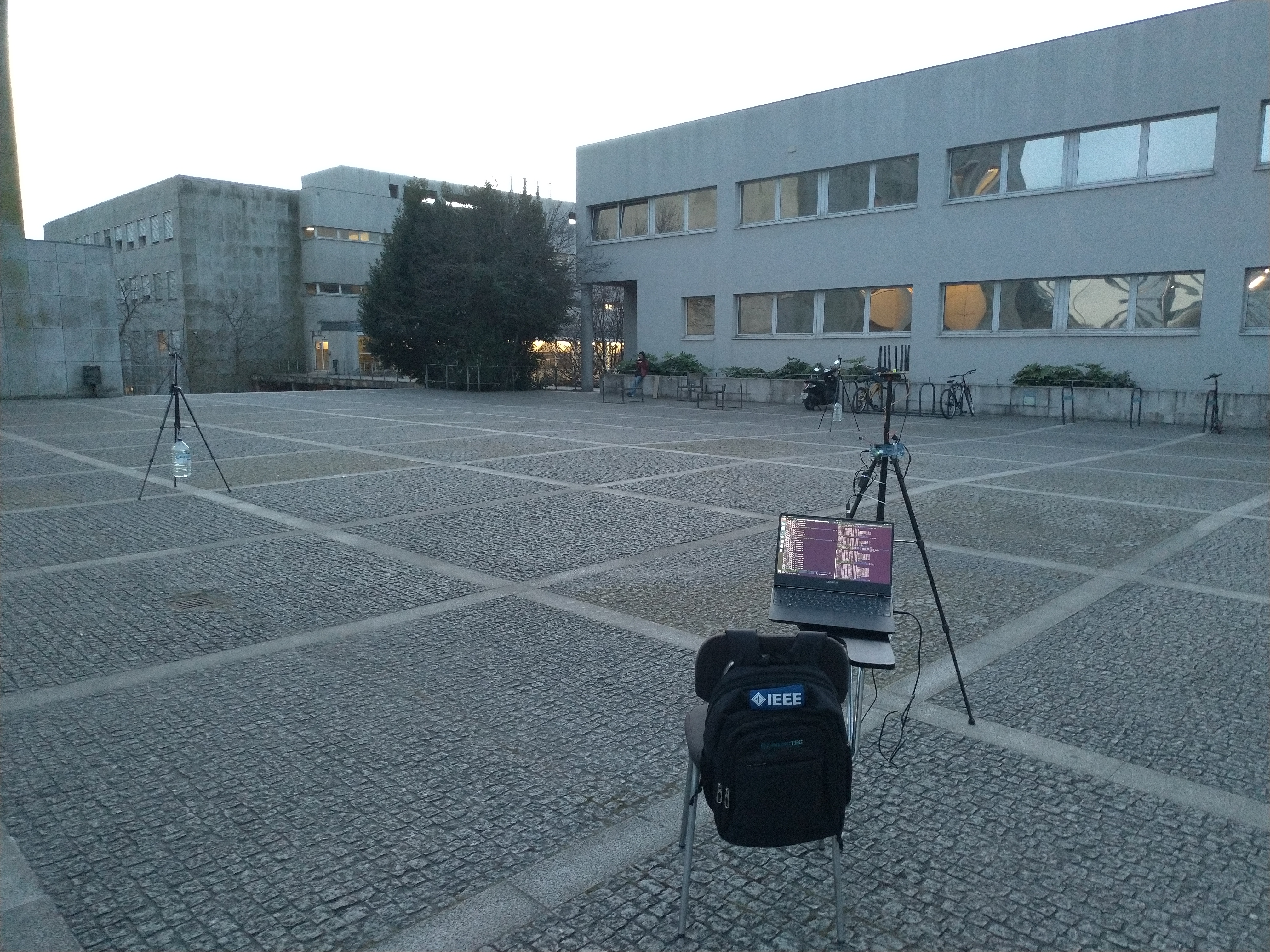}}
    \caption{Aerial view of test location (\Cref{fig:location}), and retrieval of data for one possible receiver position (\Cref{fig:groundlocation})}
\end{figure*}

The outdoor location was a flat, concrete floored lot, without obstructions. The aerial view is illustrated in \Cref{fig:location}. The total area of the lot is approximately \SI{1.5}{\kilo\meter\squared}, however, we utilized a sub-section of this area for the experiments. The area outlined in red is a square with a side of \SI{12}{\meter} in length. At each corner, we placed one of the transmitters, and for each grid location within the area, we placed the receiver, until a total of 600 packets were received for that location. The receiver faced the same absolute direction for every placement. \Cref{fig:groundlocation} shows this described setup, with the receiver placed along on of the edges of the outlined area. The boards are held on tripods, parallel to and approximately \SI{1.6}{\meter} off the ground. All the transmitters use the same respective antenna track for transmission.

\noindent\begin{figure}%
    \centering%
        \begin{tikzpicture}%
        \begin{axis}[%
        xticklabels={
            $\Delta\theta 1$,,, 
            $\Delta\theta 4$,,, 
            $\Delta\theta 7$,,,
            $\Delta\theta 10$,,, 
            $\Delta\theta 13$,,,
            $\Delta\theta 16$},
        xtick=data,
        %xticklabel style={rotate=45},
        %height=0.3\linewidth,
        height=0.5\linewidth,
        width=1\linewidth,
        grid = both,
        major grid style = {lightgray},
        minor grid style = {lightgray!25},
        xlabel=Difference Index,
        ylabel=Phase Difference,
        legend columns = 1,
        legend style={
            fill=none, 
            draw=none, 
            at={(0.45,1.4)}, 
            anchor=north,
            /tikz/every even column/.append style={column sep=0.1cm}
        }]
        
        \addplot[black, thick] 
        table [x=d, y=mean, col sep=comma] {./data/16/alldata_x2y2_board2_16.csv};
        \addlegendentry{Mean phase differences for all packets}
        
        \addplot[red, very thick] 
        table [x=d, y=filtered, col sep=comma] {./data/16/alldata_x2y2_board2_16.csv};
        \addlegendentry{Mean phase differences for packets post-filtering}

        \foreach \yindex in {2,...,169}
           \addplot[gray, only marks, mark size = 0.8]
            table [x=d, y index = \yindex, col sep=comma] {./data/16/alldata_x2y2_board2_16.csv};

        \end{axis}
        \end{tikzpicture}
\caption{Superposition of all computed phase differences (in light gray) for all received packets sent by board \#2 (\nota{a total of 170}) at the center of the test area (x=2, y=2). The average of all profiles is shown in black. Despite the variation between packets, the sinusoidal behaviour of the phase difference profile is observable. \nota{However, taking only the most frequently ocurring value for each $\Delta\theta$, a much clearer profile is achievable, shown in red.}}
\label{fig:fieldTestCenterBoard2}
% TODO: add figure from lmp code which shows how it looks with only the non reflected components

% Superposition of all computed phase differences (in light gray), however, this time we consider only the packets which, through the histogram analysis, are revealed to represent the desired (primary) packet, i.e., without any reflections or interferences from other transmissions

\end{figure}
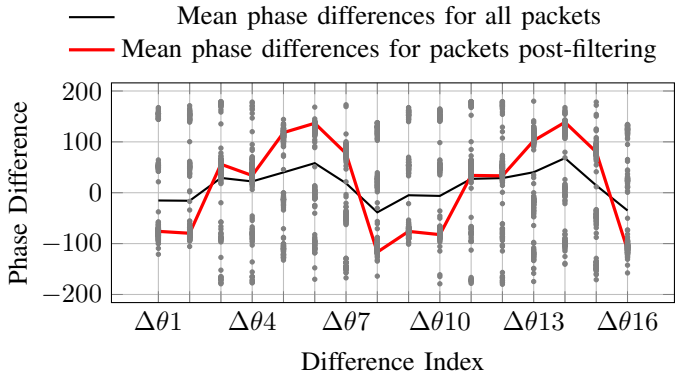

The purpose of gathering a large number of packets per transmitter, per location of the map, was to validate the expected behaviour, in a real-world setting. That is, we expected that, for a given transmitter and receiver pair, that all the packets from that transmitter, when processed, would produce phase difference profiles similar to those shown in \Cref{fig:chamberresults}. 

For brevity and clarity, we show these results for the central position of the map, only for the packets sent by board \#2, which is located at the lower right corner of the area shown in \Cref{fig:location}, from which a total of 170 packets were received. \Cref{fig:fieldTestCenterBoard2} illustrates, in light grey \nota{marks, the phase difference values between antenna pairs, e.g., $\Delta\theta1$.} 

\nota{The average phase difference profile is shown in black. The expected sinusoidal profile is observed, although, unlike the coherent behaviour observed in the anechoic chamber, the resulting profile is degraded. However, it is noticeable that, per phase difference, there are multiple clusters of points. For example, for $\Delta\theta1$, there are three noticeable groups of points. Each group represents one possible value for the phase difference $\Delta\theta1$.}

\noindent\begin{figure}%
    \centering%
        \begin{tikzpicture}%
        \begin{axis}[
            ybar,
            ymin = 0,
            ymax = 80,
            xmin = -200,
            xmax = 200,
            height=0.5\linewidth,
            width=0.96\linewidth,
            grid = both,
            major grid style = {lightgray},
            minor grid style = {lightgray!25},
            xlabel={Phase Difference Value},
            ylabel={Value Count},
            legend columns = 4,
            legend style={
            fill=none, 
            draw=none, 
            at={(0.2, 0.93)}, 
            anchor=north,
            /tikz/every even column/.append style={column sep=0.1cm}
        }]
        \addplot table [x=value, y=count, col sep=comma]{./data/testbarDelta5.csv};
        \addlegendentry{$\Delta\theta5$}
        
        \addplot table [x=value, y=count, col sep=comma]{./data/testbarDelta8.csv};
        \addlegendentry{$\Delta\theta8$}
        
        \end{axis}
     \end{tikzpicture}
\caption{\nota{Two example histograms (for 16 bins) of the values of the measured phase differences for $\Delta\theta5$ and $\Delta\theta8$, demonstrating the more frequent occurrence of the phase difference values.}}
\label{fig:histogram}     
\end{figure}
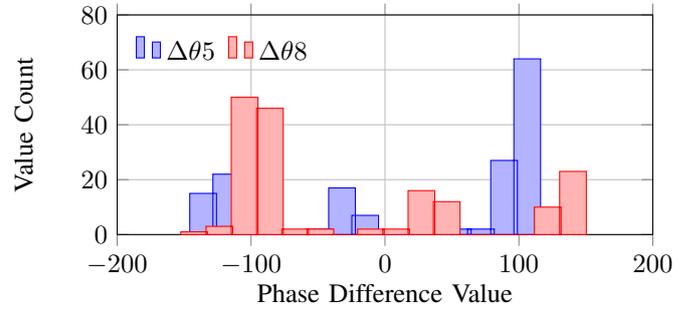
\nota{Given this, we compute one histogram per phase difference value, and determine the most frequently occurring value for that difference. This results in the average phase difference profile shown in red in \Cref{fig:fieldTestCenterBoard2}}. 
\nota{\Cref{fig:histogram} illustrates this for two of the phase difference values. The \emph{x} axis shows the possible values for the phase difference, while the \emph{y} axis shows the counts for the value ranges within the 16 bins. For both examples, there is an observable higher count for two values, while two other frequent values occur, at much lower counts.}

\noindent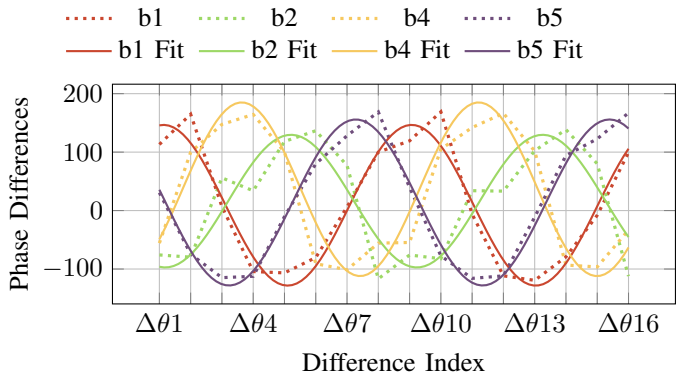
\begin{figure}%
    \centering%
        \begin{tikzpicture}%
        \begin{axis}[%
        xticklabels={
            $\Delta\theta 1$,,, 
            $\Delta\theta 4$,,, 
            $\Delta\theta 7$,,,
            $\Delta\theta 10$,,, 
            $\Delta\theta 13$,,,
            $\Delta\theta 16$},
        xtick=data,
        %xticklabel style={rotate=45},
        %height=0.3\linewidth,
        height=0.5\linewidth,
        width=1\linewidth,
        grid = both,
        major grid style = {lightgray},
        minor grid style = {lightgray!25},
        xlabel=Difference Index,
        ylabel=Phase Differences,
        legend columns = 4,
        legend style={
            fill=none, 
            draw=none, 
            at={(0.38,1.4)}, 
            anchor=north,
            /tikz/every even column/.append style={column sep=0.1cm}
        }]

        \addplot[chart1, dotted, very thick]
        table [x=d, y=b1, col sep=comma]
        {./data/16/filered_board_packets_mean_at_x2y2.csv};
        \addlegendentry{b1}
        
        \addplot[chart2, dotted, very thick]
        table [x=d, y=b2, col sep=comma]
        {./data/16/filered_board_packets_mean_at_x2y2.csv};
        \addlegendentry{b2}
        
        \addplot[chart4, dotted, very thick]
        table [x=d, y=b4, col sep=comma]
        {./data/16/filered_board_packets_mean_at_x2y2.csv};
        \addlegendentry{b4}
        
        \addplot[chart3, dotted, very thick]
        table [x=d, y=b5, col sep=comma]
        {./data/16/filered_board_packets_mean_at_x2y2.csv};
        \addlegendentry{b5}
        
        % add the sine fit as a thick wave over the dotted waves!
        \addplot[chart1, thick]
        table [x=d, y=b1, col sep=comma]
        {./data/16/filered_board_packets_mean_at_x2y2_sineFits.csv};
        \addlegendentry{b1 Fit}
        
        \addplot[chart2, thick]
        table [x=d, y=b2, col sep=comma]
        {./data/16/filered_board_packets_mean_at_x2y2_sineFits.csv};
        \addlegendentry{b2 Fit}
        
        \addplot[chart4, thick]
        table [x=d, y=b4, col sep=comma]
        {./data/16/filered_board_packets_mean_at_x2y2_sineFits.csv};
        \addlegendentry{b4 Fit}
        
        \addplot[chart3, thick]
        table [x=d, y=b5, col sep=comma]
        {./data/16/filered_board_packets_mean_at_x2y2_sineFits.csv};
        \addlegendentry{b5 Fit}
        
        \end{axis}
        \end{tikzpicture}
\caption{\nota{Phase difference profiles between the receiver, place at the center of the test area, and all beacons. In dotted lines we show the phase differences after histogram based filtering, and in solid lines a sinusoidal fit. The four waves display phase differences of \SI{90}{\degree}, e.g., between \emph{b1}  and \emph{b4}, which is the expected behaviour given the physical disposition of the boards}}
\label{fig:centerfiltered}
\end{figure}

Finally, we can demonstrate the relationship between the horizontal displacement of the phase difference profile and the \ac{AoA}, as shown in \Cref{fig:centerfiltered}. By placing the receiver at the center of the map (always with the same absolute orientation), the \acp{AoA} received from any two neighbouring beacons (e.g., \emph{b1} and \emph{b5}) will be offset by plus or minus \SI{90}{\degree}. 
By defining one specifc phase difference profile as a reference for the receiver's relative \SI{0}{\degree}, we can attain the \ac{AoA} by determining the displacement relative to that reference. 

%%%%%%%%%%%%%%%%%%%%%%%%%%%%%%%%%%%%%%%%%%%%%%%%%%%%%%%%%%%%%%%%%%%%%%%%%%%
\section{Conclusion}

This paper has presented a design of a circular 8-antenna array, with the intent of designing a self-localizing receiver. We briefly explained the methodology of phase difference calculation between neighbouring antenna pairs, and how the eight differences can be combined to estimate an \ac{AoA}. 

We have demonstrated the effect that noise in the raw phase measurements can produce in the phase difference profile via simulation, and conducted two experimental evaluations. The first demonstrated the correct behaviour of the design versus the theoretical model in an anechoic chamber, as a function of the relative orientation of the receiver and transmitter. The second was a field test using four transmitters placed at the vertices of a \SI{12}{\meter} by \SI{12}{\meter} area. 
%In this environment, 
Although it is possible to extract the expected phase difference profiles, the receiver must remain still at the same location, so that the effects of noise can be averaged out by reception of multiple packets. 
%
% We find that in a real-world environment, the sampled phase data is significantly affected by noise, meaning that solutions for self-locating edge devices with limited computing power requires further study, as computationally demanding algorithms are difficult to implement...
% NEW WORK...
% historigram based reflection removal
\nota{Regarding pre- processing, we have also briefly demonstrated how reflections may be filtered out and attain the desired behaviour. We will attempt to explore this further, by resorting to the currently unused magnitude values and the best known location of the receiver itself.}

In order to allow for the capability of self-localization on this kind of embedded device with limited computing power, subject to movement and relying on the small number of packets, as we have previously evaluated \cite{9482525}, requires further study, as computationally demanding algorithms may compromise how quickly a position can be computed, especially since multiple \acp{AoA} are required. 

% ... as employment of computationally complex algorithms may compromise how quickly a position can be computed, especially since multiple \acp{AoA} are required. 

As it is difficult to reduce the error from the obtained phase measurements themselves, future work will focus on extracting more robust phase differences despite this noise. We plan to evaluate the viability of machine learning models to extract the \ac{AoA} from phase difference data, as some recent approaches are attempting \cite{8554304,9336252}, especially to reduce the computational cost of the existing algorithms \cite{1143830}.

% TODO: at the end,push the discussion towards custom edge devices to fulfill the real time compute of everything, which ties with the argument of edge compute etc

% TODO: the conclusion should mention concerns as the reports, such as, would the 8 samples make the result better? would batching packets produce better results by decreasing TX time via uart? how much of the computation can be moved onto the embedded microcontroller? can hwardware acceleration or a custom system solve the problem?

%%%%%%%%%%%%%%%%%%%%%%%%%%%%%%%%%%%%%%%%%%%%%%%%%%%%%%%%%%%%%%%%%
%\section*{Acknowledgment}

%This work is financed by the ERDF – European Regional Development Fund through the Operational Programme for Competitiveness and Internationalisation - COMPETE 2020 Programme within project SLID: Stock Live IDentification (POCI-01-0247-FEDER-045388).

\bibliographystyle{IEEEtran}
\bibliography{main}
\end{document}